\documentclass[prb,showpacs,preprintnumbers,amsmath,amssymb,twocolumn,superscriptaddress,longbibliography]{revtex4-1}
\usepackage{graphicx}
\def\be{\begin{equation}}
\def\ee{\end{equation}}
\def\bea{\begin{eqnarray}}
\def\eea{\end{eqnarray}}
\begin{document}
\def\non{\nonumber}
\title{Commensurate to incommensurate magnetic phase transition in Honeycomb-lattice pyrovanadate Mn$_2$V$_2$O$_7$}
\author{J. Sannigrahi}
\email{jhuma.iacs@gmail.com} 
\affiliation{ISIS Facility, Rutherford Appleton Laboratory, STFC, Chilton, Didcot, OX11 0QX, United Kingdom}
\author{D.T. Adroja}
\email{devashibhai.adroja@stfc.ac.uk}
\affiliation{ISIS Facility, Rutherford Appleton Laboratory, STFC, Chilton, Didcot, OX11 0QX, United Kingdom}
\affiliation{Highly Correlated Matter Research Group, Physics Department, University of Johannesburg, P.O. Box 524, Auckland Park 2006, South Africa}
\author{R. Perry}
\author{M. J. Gutmann}
\affiliation{ISIS Facility, Rutherford Appleton Laboratory, STFC, Chilton, Didcot, OX11 0QX, United Kingdom}
\author{V. Petricek}
\affiliation{Institute of Physics, Czechoslovak Academy of Sciences, Na Slovance 2, 18040 Praha 8, Czechoslovakia}
\author{D. Khalyavin}
\affiliation{ISIS Facility, Rutherford Appleton Laboratory, STFC, Chilton, Didcot, OX11 0QX, United Kingdom}

\pacs {75.25.+z, 75.50.-y, 75.50.Ee,  75.30.Kz, 74.62.-c, 75.47.-m, 71.70.Gm}

\date{\today}

\begin{abstract}
We have synthesized single crystalline sample of Mn$_2$V$_2$O$_7$ using floating zone technique and investigated the ground state using magnetic susceptibility, heat capacity and neutron diffraction. Our magnetic susceptibility and heat capacity reveal two successive magnetic transitions at $T_{N1} =$ 19 K and $T_{N2} =$ 11.8 K indicating two distinct magnetically ordered phases. The single crystal neutron diffraction study shows that in the temperature ($T$) range 11.8 K $\le T \le$ 19 K the magnetic structure is commensurate with propagation vector $k_1 = (0, 0.5, 0)$, while upon lowering temperature below $T_{N2} =$ 11.8 K an incommensurate magnetic order emerges with $k_2 = (0.38, 0.48, 0.5)$ and the magnetic structure can be represented by cycloidal modulation of the Mn spin in $ac-$plane. We are reporting this commensurate to incommensurate transition for the first time. We discuss the role of the magnetic exchange interactions and spin-orbital coupling on the stability of the observed magnetic phase transitions.

\end{abstract} 
\maketitle
\section{Introduction}
Bulk crystals are inherently three dimensional; however, they may consist of magnetic ions whose spins interact only along one or two certain crystallographic directions. Such type of compounds are known as low dimensional magnets which have aroused considerable attention in solid state chemistry and physics due to their non-typical behaviour which mainly deal with electrical and magnetic properties. Two dimensional honeycomb-lattice systems are one of the interesting low dimensional magnetic systems which have been intensively investigated because of novel ground states induced by frustration and quantum fluctuations.~\cite{meng, ladak, nakatsuji, grandi, zheng} Magnetic order can be obtained on a regular two dimensional lattice assuming antiferromagnetic interactions between adjacent spins;~\cite{wenzel} or on a bilayer lattice with frustrating interlayer interactions.~\cite{tutsch} In recent times, these honeycomb lattice systems have been extensively investigated both theoretically and experimentally as potential candidate for Kitaev quantum spin liquid state, for example Na$_2$IrO$_3$,  $\alpha$-RuCl$_3$ .~\cite{yan, nishimoto} Moreover, superconductivity has also been observed in pnictide SrPtAs with honeycomb lattice.~\cite{youn}

\par
The title compound Mn$_2$V$_2$O$_7$ is a member of the family of transition metal-vanadium oxides M$_2$V$_2$O$_7$ (M $=$ Cu, Ni, Co, Mn) which have attracted much interest due to their rich structural features and magnetic properties.~\cite{tou, he1, tsirlin, he2, he3, calvo, jhuma1, jhuma2, jhuma3} Mn$_2$V$_2$O$_7$ is composed of magnetic Mn$^{2+}$ (3$d^5$, $S = 5/2$) and non-magnetic V$^{5+}$ (3$d^0$, $S = 0$) ions and was reported to possess quasi two dimensional distorted honeycomb lattice.~\cite{liao, he4} It exhibits two different structural phases; namely $\beta-$ phase (above $\approx$ 310 K) with monoclinic ($C2/m$) symmetry and $\alpha-$ phase (below $\approx$ 284 K) with triclinic ($P\bar{1}$) symmetry. As temperature decreases, the high temperature monoclinic symmetry of Mn$_2$V$_2$O$_7$ reduces to triclinic one and this structural transition is completely reversible with reasonable thermal hysteresis, in line with the martensitic-like first order nature of this structural transition involving lattice distortion but without atomic exchange.~\cite{jhuma2, he4} Depending on the synthesis procedure these transition temperatures vary because of the change in O$_2$ pressure and cooling rate.  The honeycomb networks of Mn-atoms in the $\beta$ phase are parallel to (001) plane and those in the $\alpha$ phase are parallel to the (110) plane. Further the sample undergoes paramagnetic to antiferromagnetic transition below around 19 K. It is noted that such properties of Mn$_2$V$_2$O$_7$ were mainly based on macroscopic characterization methods on polycrystalline as well as flux grown single crystalline samples. But a detailed microscopic model of the magnetism has not yet been developed.

\par
So far there are few reports on the single crystal growth of Mn$_2$V$_2$O$_7$ where only flux growth technique has been followed.~\cite{he5, zhou} The crystals obtained are small and moreover the grown crystals may incorporate traces of the molten flux or the crucible materials, which is highly undesirable. Keeping this view in mind, we decided to grow Mn$_2$V$_2$O$_7$ crystals using the traveling solvent floating zone (TSFZ) method associated with an optical image furnace. Mn$_2$V$_2$O$_7$ melts congruently upon heating at (1080 $\pm$ 3)$^{\circ}$ C.~\cite{fotiev} In order to understand the structural and magnetic properties of the ground state of Mn$_2$V$_2$O$_7$ at the macroscopic as well as microscopic level, we carried out x-ray diffraction (XRD), magnetization and heat capacity measurement along with the single crystal neutron diffraction study. 

\begin{figure}[t]
\centering
\includegraphics[width = 8 cm]{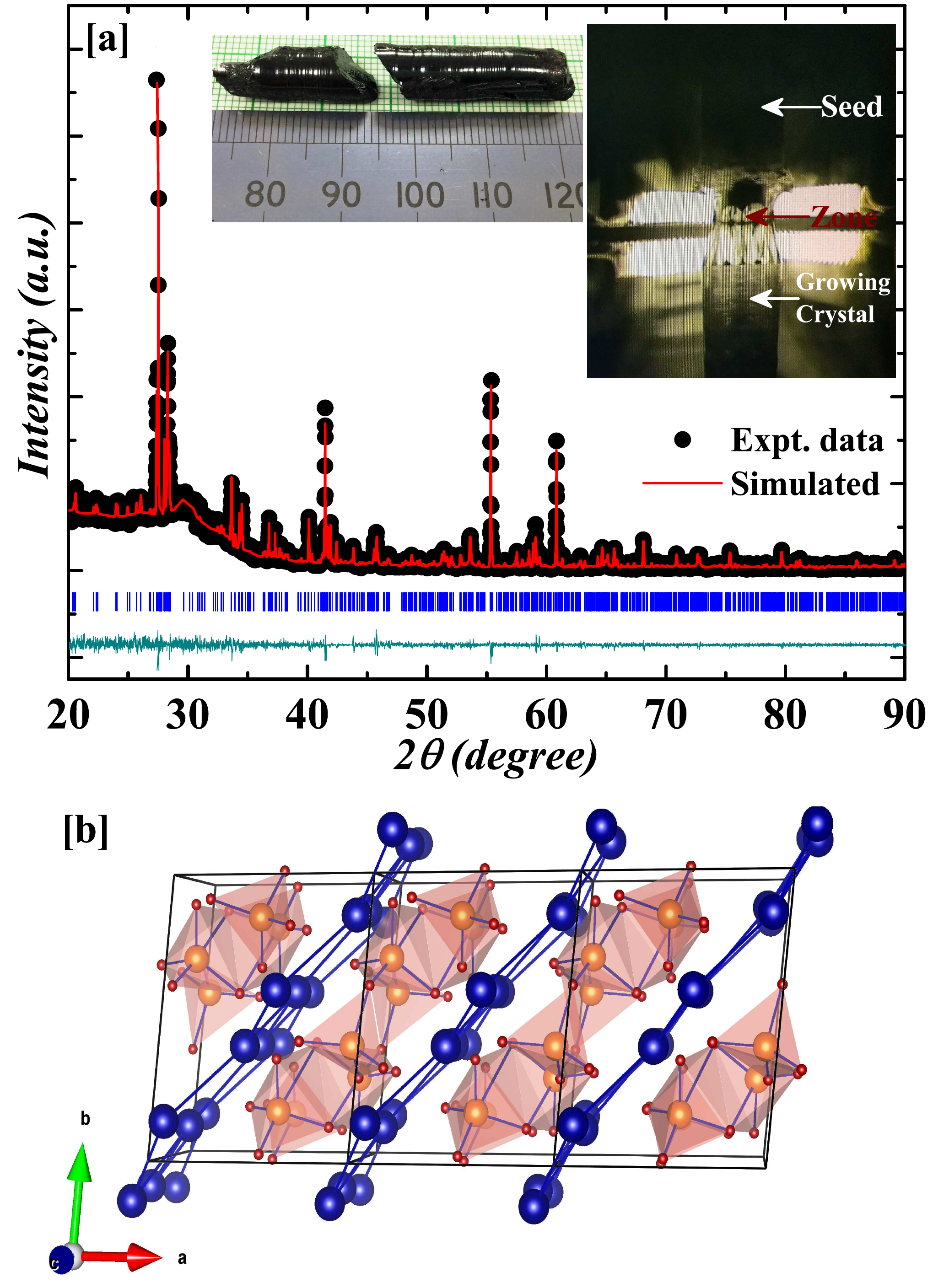}
\caption{(a) Main panel shows the XRD pattern of crushed single crystal measured at $T =$ 100 K. Black scattered points and red solid line represent respectively the experimental data and the simulated curve obtained from Rietveld refinement. Blue ticks indicate the Bragg positions for triclinic $P\bar{1}$ structure and the cyan line shows the difference between experimental and calculated pattern. Right inset shows the stable solvent zone formation during growing, while left inset depicts the picture of the as grown crystal. (b) depicts the perspective view of the crystal structure of Mn$_2$V$_2$O$_7$ at 100 K where blue, yellow and red spheres represent Mn, V and O atoms respectively.}
\label{XRD}
\end{figure}

\section{Experimental details}
The single crystalline sample of Mn$_2$V$_2$O$_7$ has been grown in a four-mirror optical floating zone furnace from Crystal System Corporation, Japan. The furnace consists of four halogen lamps. Every lamp takes the focus position of an ellipsoidal reflector whereas the second focus of the reflectors coincide at a point inside growth chamber on the vertical axis of the furnace. A molten zone between the polycrystalline feed rod and the seed rod is formed at this common focal point where the energy flux of the lamps converge after reflecting from the reflectors. The feed rod is suspended from the upper shaft with a nickel wire while the seed rod is clamped to the lower shaft using an alumina holder. Polycrystalline sample of Mn$_2$V$_2$O$_7$ was prepared by standard solid state reaction route in air. Highly pure MnO$_2$ and V$_2$O$_5$ were mixed thoroughly in a stoichiometric ratio in an agate mortar.The mixture was pressed into pellets and sintered at 600$^{\circ}$C for one week with several intermediate grindings. The phase analysis of the final product was performed using powder x-ray diffraction (Miniflex). The single phase powder of Mn$_2$V$_2$O$_7$ was filled in rubber tube and cold pressed under 700 bar isostatic pressure to obtain uniform rod. The rods formed in this process are 5 mm in diameter and 70 - 90 mm in length. Then the rods were sintered in the platinum boat  at the temperature 900$^{\circ}$C overnight to obtain very dense and homogeneous rod. This process ensures the homogeneous melting of the feed rod during growth process. A part about 20 mm long was cut from the rod and used as the seed rod.

\par
 Three different growth experiments were performed with different oxygen pressure and growth speed employed; (i) 60$\%$ O$_2$ $+$ 40$\%$ Ar and growth speed 5 mm/hr (ii) 100 $\%$ O$_2$ flow ($P =$ 7 bar) and growth speed 5 mm/hr and (iii) 100 $\%$ O$_2$ flow ($P =$ 7 bar) and growth speed 1 mm/hr. The growth process is initiated by fusing the bottom end of the feed rod inside the optical furnace. In our case 300 watt lamps have been installed. The upper and lower shafts of the furnace were rotated in opposite directions with rotation speed 30 rpm to improve the temperature and compositional homogeneity of the float zone. In the first two attempts, upper shaft was translated at a speed of 1 mm/hr and in the third attempt it is translated at a speed of 0.5 mm/hr in order to control the steady state of floating zone. In the first attempt molten zone was not fully stable and formation of bubbles were observed near the upper edge of the zone possibly due to the reduction of Mn ions. Then we decided to use 100 $\%$ O$_2$ pressure to prevent the reduction of Mn$^{2+}$ and with this environment we got significantly stable zone throughout the growth process as shown in the right inset of figure 1(a).  

\par
The grown crystals were characterized using powder x-ray diffraction of crushed crystal (Bruker D8 Advance), Laue x-ray diffraction and single crystal neutron diffraction in SXD diffractometer, ISIS neutron and muon facility. The magnetization of the crystals was measured by applying an external field along the three different crystallographic directions using a vibrating sample magnetometer (SQUID VSM, Quantum Design) in the temperature range $T =$ 2 to 320 K and applied magnetic field ($H$) ranging from -70 to +70 kOe. Heat capacity has been measured in Quantum Design PPMS. The x-ray diffraction (XRD) pattern of crushed crystal measured at 100 K is shown in the main panel of figure 1(a). Rietveld refinement on powder XRD data was performed using Fullprof software package and the data is well fitted using triclinic structure with $P\bar{1}$ space group. The crystal structure (shown in figure 1(b)) is consistent with the previously reported literature. 

\begin{figure}[t]
\centering
\includegraphics[width = 8 cm]{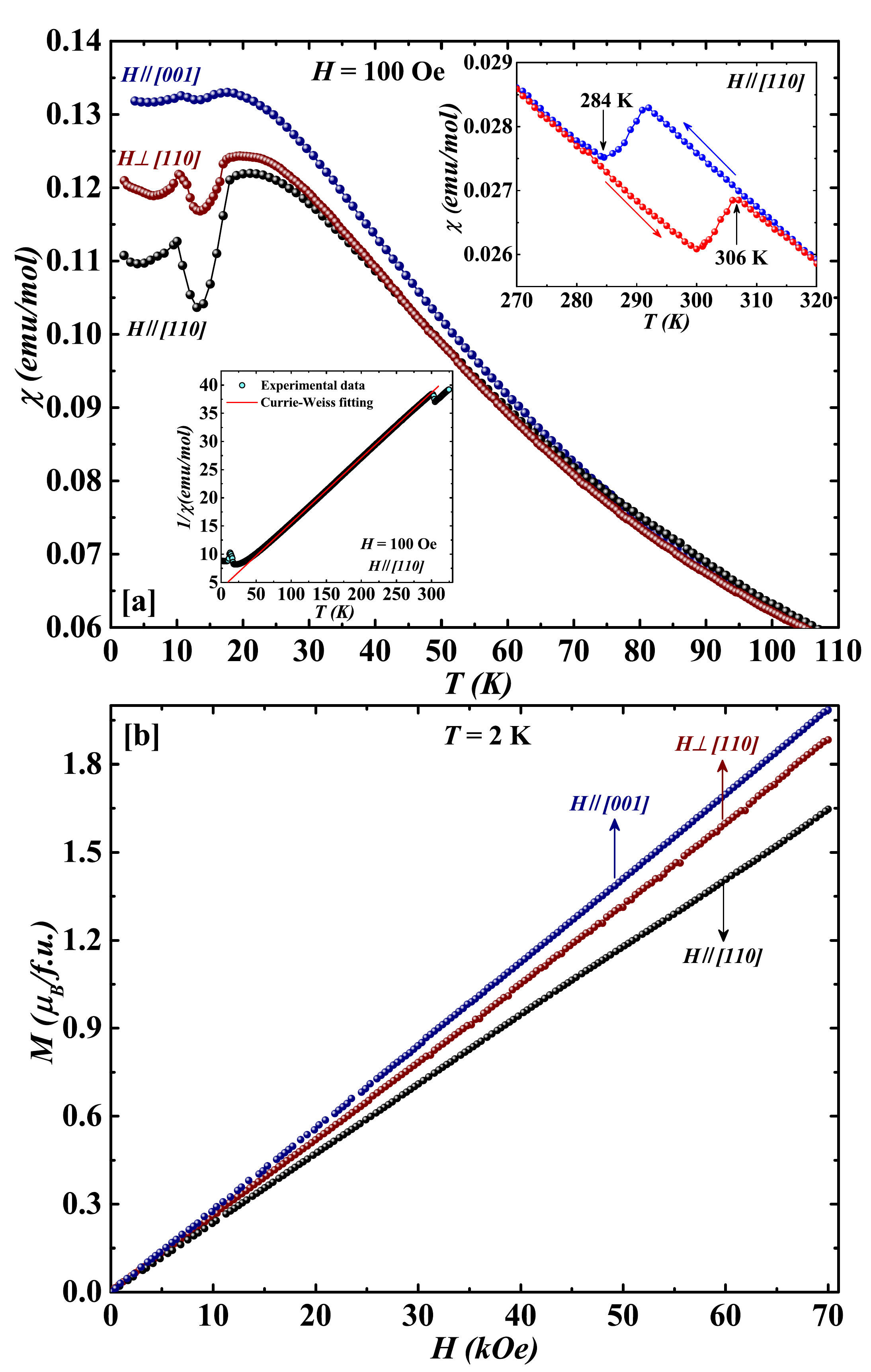}
\caption{(a) Main panel shows the temperature dependent magnetization data in the zero field-cooled heating protocol where magnetic field applied along three crystallographic directions: $H \parallel (110)$, $H \perp (110)$ and $H \parallel (001)$. Upper inset depicts the thermal hysteresis observed between 284 K and 306 K on field cooling and field-cooled heating measurements whereas lower inset represents the Curie-Weiss fitting in the paramagnetic region. (b) $M$ versus $H$ curves are displayed measured at $T =$ 2 K along three different crystallographic directions.}
\label{magnetization}
\end{figure}

\begin{figure}
\centering
\includegraphics[width = 8 cm]{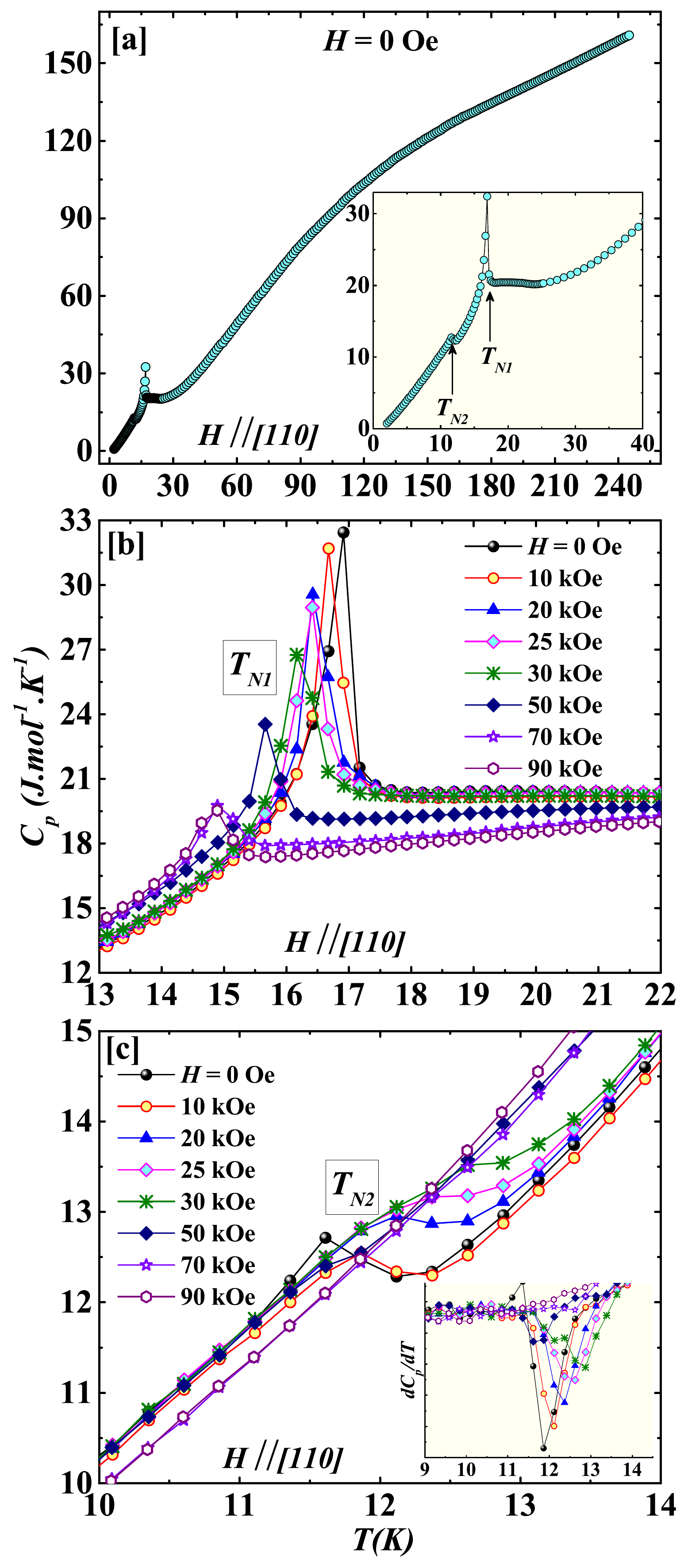}
\caption{(a) Main panel shows the temperature dependent heat capacity measured at zero applied magnetic field while inset shows the enlarged view of the low-$T$ regime indicating the two transitions at $T_{N1}$ and $T_{N2}$. The magnetic field dependence of $T_{N1}$ and $T_{N2}$ are depicted in (b) and (c) respectively. $dC_p/dT$ vs $T$ is plotted in the inset of (c) to show clearly the change in $T_{N2}$ with $H$.}
\label{HC}
\end{figure}

\section{Magnetization and heat capacity}
The main panel of figure 2(a) shows the temperature dependence of magnetic susceptibility ($\chi$) from 2 K to 110 K, which was measured in zero field cooled heating protocol under an applied magnetic field of $H =$ 100 Oe along the parallel and perpendicular to (110) plane and parallel to (001) plane of $\alpha-$Mn$_2$V$_2$O$_7$. A clear signature of antiferromagnetic (AFM) transition is observed at $T_{N1} \approx$ 19 K which matches quite well with the previously reported data.~\cite{he4, ueda, liao} Additionally we found another anomaly at around $T_{N2} =$ 11.8 K in all three crystallographic directions. Although the $T_{N1}$ and $T_{N2}$ are present at the same temperatures, but different histories are clearly observed below 19 K dependent on direction of the applied magnetic field. This indicates the presence of magnetic anisotropy in this system. Upper inset of figure 2(a) shows the thermal hysteresis within 284 K and 306 K between field-cooling and field-cooled-heating protocols of measurement. The existence of hysteresis suggests that this reversible structural transition is a first order martensitic-like transition where thermo-elastic solid-solid phase transition occurs involving lattice distortion but without atomic exchange. The $1/\chi(T)$ curve follows the Curie-Weiss law above 50 K in the $\alpha-$ phase of the compound as shown in the lower inset of figure 2(a). We obtained Curie-Weiss $T$, $\theta_C =$ -34 K indicating the predominant AFM interaction and an effective paramagnetic moment, $\mu_{eff} =$ 5.86 $\mu_B$/Mn$^{2+}$, which is nearer to the spin-only value  5.92 $\mu_B$ of Mn$^{2+}$. Figure 2(b) shows the isothermal magnetization measured at $T =$ 2 K under applied magnetic field up to 70 kOe where $H \parallel (110)$, $H \perp (110)$ and $H \parallel (001)$. Linear nature of the $M$ versus $H$ curves in all three directions indicates again the presence of predominant AFM interaction in the studied compound. From the magnetic susceptibility and magnetization isotherm measurements it is clear that the $c$-axis is an easy magnetization axis with weak anisotropy.  

\par
The heat capacity ($C_p$) as a function of temperature measured at zero applied magnetic field is depicted in figure 3(a). At high temperature, $C_p$ is completely dominated by phonon excitations. At low temperature, two $\lambda-$type anomalies are observed at $T_N \approx$ 19 K and $T_{N2} \approx$ 11.8 K indicating two magnetic phase transitions consisting with the magnetization measurement. To gain more information about magnetic ordering, $C_p(T)$ has been measured in different applied magnetic field from 0 Oe to 90 kOe which is shown in figure 3(b) and (c). With increasing $H$, $T_{N1}$ shift towards lower temperature suggesting antiferromagnetic nature of this transition. In addition to this, the intensities of the peaks are decreasing with increasing magnetic field because of the redistribution of magnetic entropy. On the other hand, the peak at $T_{N2}$ shifts slightly towards higher temperature from $H =$ 0 Oe to 30 kOe up to a maximum $T_{N2} \approx$ 12.8 K and decreases more clearly with further increase of $H$. This peak also becomes broader with increasing $H$ and finally completely suppressed above 50 kOe.

\begin{figure}
\centering
\includegraphics[width = 5.7 cm, trim=0 0 -80 -60, clip]{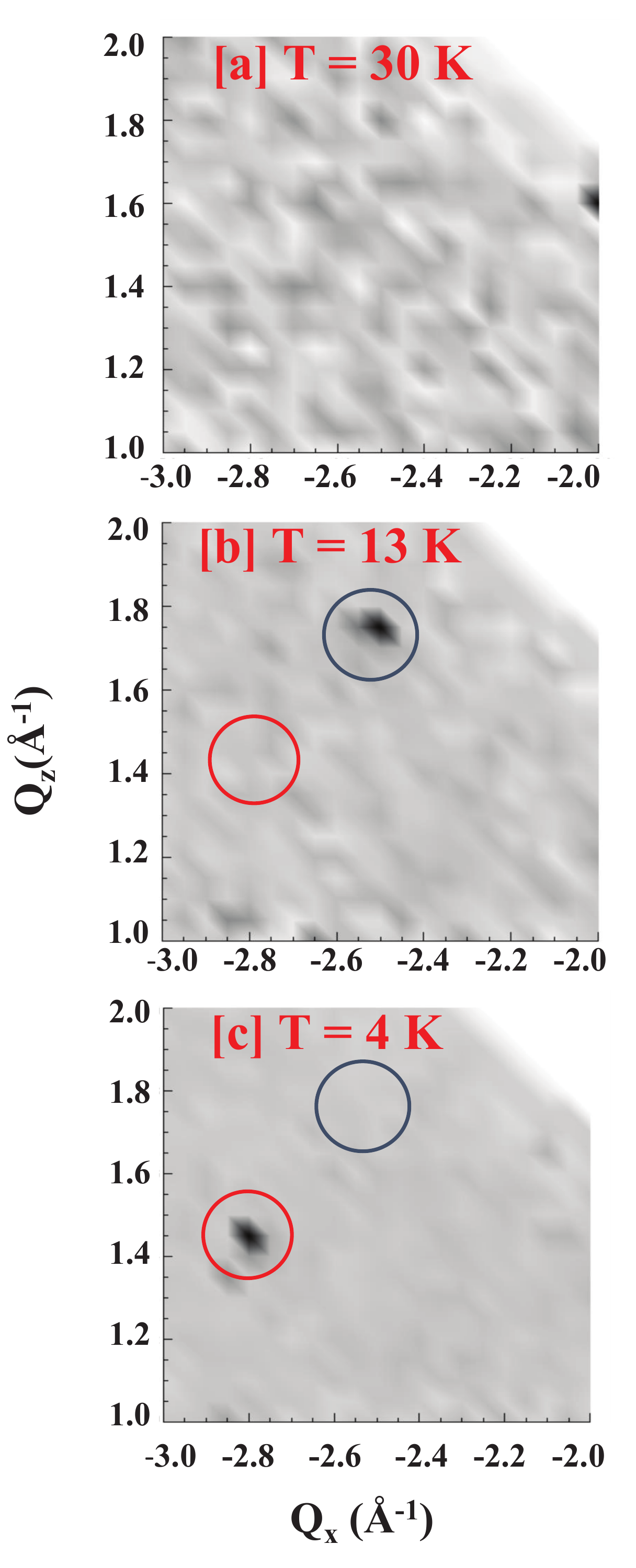}
\caption{2D intensity maps of the reciprocal $(h,0.5,l)$ planes at 30 K, 13 K and 4 K are depicted in (a), (b) and (c) respectively. The peak for commensurate propagation vector $k_1 =$ (0,0.5,0) at 13 K is shown by blue circle in figure(b) while the magnetic peak for incommensurate propagation vector $k_2 =$ (0.38,0.48,0.5) at 4 K is indicated by red circle in figure(c).}
\label{SXD}
\end{figure}

\section{Single Crystal Neutron Diffraction}
Single-crystal neutron diffraction study was carried out on the SXD diffractometer at the ISIS Neutron and Muon facility which utilizes the time-of-flight Laue technique.~\cite{keen} A Mn$_2$V$_2$O$_7$ single crystal of approximate dimensions 2 $\times$ 1 $\times$ 0.5 mm$^3$ ($m \approx$ 50 mg) was mounted at the end of an Aluminium pin using the adhesive Al tape. For reliable determination of the propagation vector single crystal neutron diffraction data were collected at different temperatures ranging from 4 K to 30 K with special attention at temperatures $T =$ 4, 12, 13 and 30 K using a closed cycle He refrigerator. Five exposures in different crystal orientations with respect to incident neutron beam were collected for 1 hr each. The data were processed using locally available SXD2001 software.~\cite{SXD2001} Structural refinements of the neutron data were performed with the JANA 2006~\cite{JANA} software using standard scattering length densities and the Mn$^{2+}$ form factor. The neutron intensity in a part of the reciprocal ($h,1/2,l$) scattering plane in $\alpha$-Mn$_2$V$_2$O$_7$ is shown as two dimensional intensity maps in figure 4 for $T =$ 30 K, 13 K and 4 K. To confirm the nuclear structure a refinement of the data set collected at 30 K (well above the peak in the magnetic susceptibility and heat capacity) has been performed which confirms the triclinic structure with the $P\bar{1}$ space group. The crystal structures parameters are consistent with the already reported in the literature. In the intermediate phase (in between $T_{N2}$ and $T_{N1}$) we observed a set of magnetic superstructure peak corresponding to the commensurate propagation vector $k_1 =$ (0,0.5,0) which are marked by blue circle in figure 4(b) at $T =$ 13 K. But in the ground state, below $T_{N2} =$ 11.8 K this commensurate phase disappears and a different set of magnetic superstructure peak arises as shown in the figure 4(c) marked with red circle. Taking the positions of these new sets of magnetic peaks, it is evident that the magnetic order at lowest temperature is incommensurate with propagation vector $k_2 =$ (0.38, 0.48, 0.5). No change in the incommensurate value of $k_2$ with temperature has been detected within the measurement accuracy.

\begin{figure}[t]
\centering
\includegraphics[width = 8 cm]{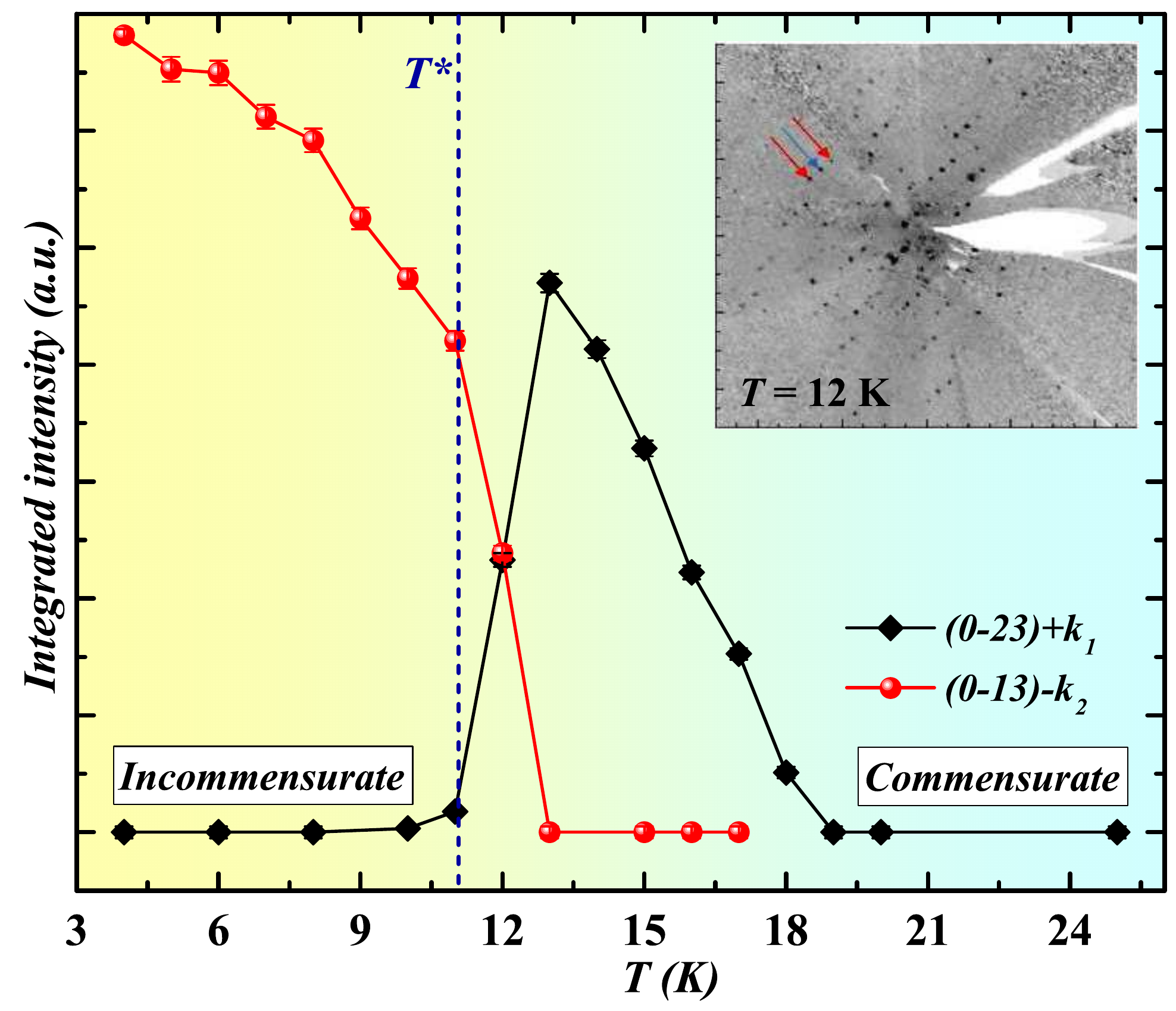}
\caption{(a) Main panel shows the temperature dependence of integrated intensity of two magnetic superstructure peaks indicating the thermal evolution of commensurate and incommensurate magnetic phases. Inset shows the two dimensional intensity map of the reciprocal $(h,0.5,l)$ at 12 K where indicating the presence of both propagation vectors $k_1$ (blue arrow) and $k_2$ (red arrow).}
\label{OP}
\end{figure}

\par
The two magnetic superstructure peaks observed in neutron diffraction pattern of Mn$_2$V$_2$O$_7$, $(0\bar{2}3)^+$ and $(0\bar{1}3)^-$ satellite peaks for commensurate and incommensurate phase respectively, have been recorded at several temperatures. Here, $(hkl)^{\pm}$ $=$ $(hkl) \pm k_1/k_2$ denotes a satellite peak of the nuclear $(hkl)$ peak. The integrated intensities of these peaks are plotted as a function of temperature as shown in the figure 5. Magnetic intensity fully vanishes above $T_{N1} =$ 19 K in close agreement with other bulk measurements. The intensity of the commensurate peak disappears below $T_{N2} =$ 11.8 K. Moreover, as displayed in the inset of figure 5, the pattern recorded at $T =$ 12 K, shows the presence of magnetic superstructure peaks of both commensurate as well as incommensurate phases which means phase separation takes place indicating a first order transition from low temperature incommensurate to commensurate intermediate temperature phase. This is consistent with the first order nature of this transition observed from heat capacity and magnetization measurements.

\begin{figure}[t]
\centering
\includegraphics[width = 8 cm]{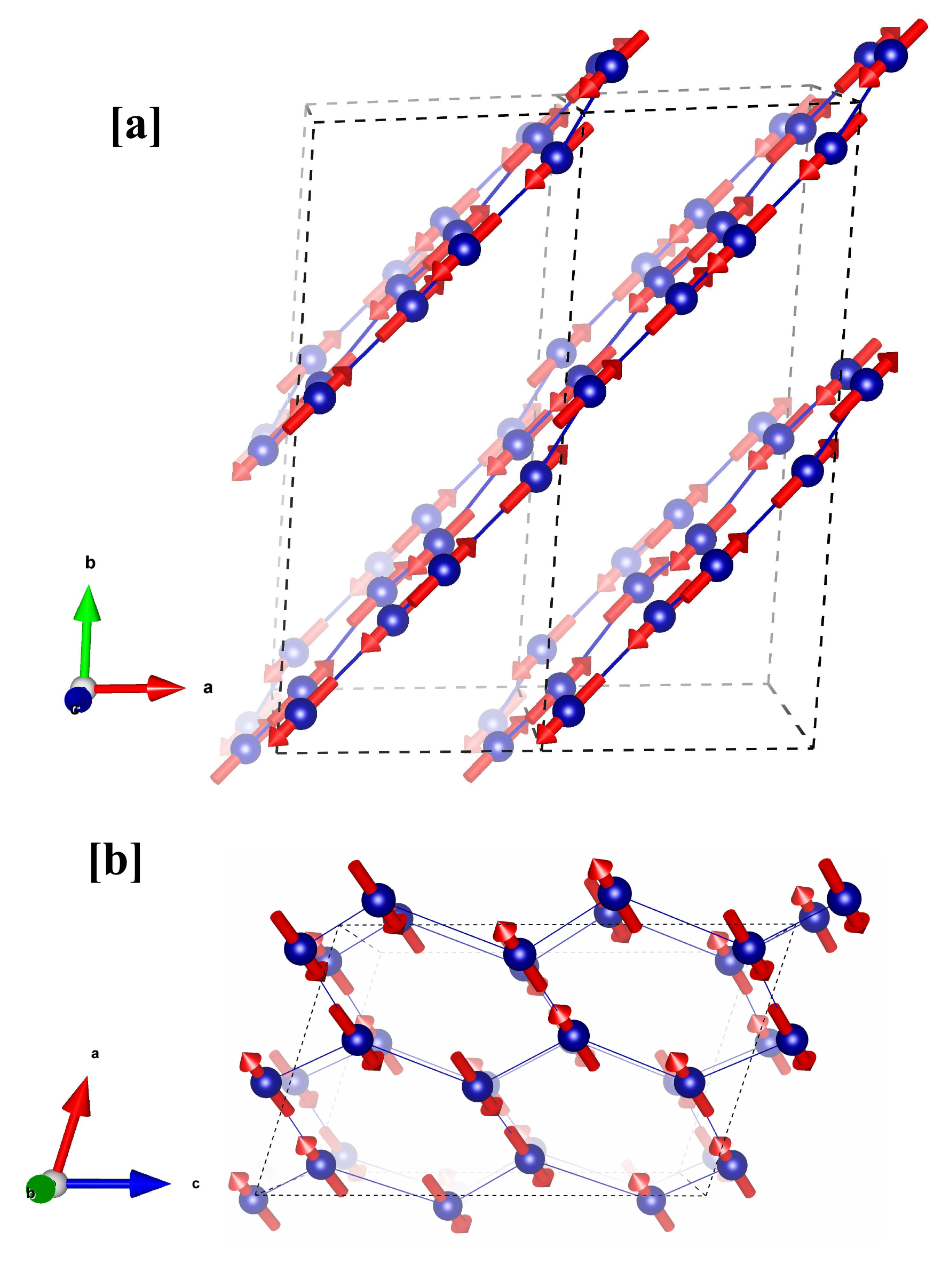}
\caption{(a) Perspective view of spin configuration in the commensurate phase ($T =$ 13 K) is shown. (b) shows the arrangement of Mn$^{2+}$ spins on the distorted honeycomb layer for the same temperature.}
\label{COM}
\end{figure}

\par
To reveal the magnetic structure, refinement of two data sets collected at 4 K and 13 K, has been performed with JANA2006. Magnetic representation analysis sets symmetry restrictions only on the high-temperature commensurate phase. The restrictions imply that the magnetic structure involves either ferromagnetic ($\Gamma_1$) or antiferromagnetic ($\Gamma_2$) coupling of the spins related by inversion in the 2i Wyckoff position. There are no restrictions on the moments direction and Mn in the symmetry independent positions can adopt different size and orientation. The refinement however was constrained to keep the structure collinear with the same ordered moment for the all four Mn sites. The obtained solution is shown in Figure 6, which corresponds to $\Gamma_2$ irreducible representation of the $P\bar{1}$ space group. Magnetic moments lie mostly in the $ac$-plane with a small component along the $b$-axis. The size of the ordered moment at $T =$ 13 K has been determined to be 2.83 $\mu_B$ per Mn. 
In the incommensurate phase, which onsets below $T_{N2} =$ 11.8 K, symmetry sets no restrictions and we tried to fit the low-temperature diffraction data considering three different models: (a) amplitude modulated structure derived from helical structure, (b) cycloidal structure where moments are rotating with respect to a plane and (c) proper helical structure. The statistically best fit is obtained for the cycloidal model with equal amplitude of magnetic moment on each Mn atom. It yields for the ordered magnetic moment of $\approx$ 5.19 $\mu_B$ per Mn. Figs. 7(a) and (b) show the proposed cycloidal magnetic structure in the low temperature incommensurate phase from the $c-$axis and honeycomb plane respectively. The rotation of the spins is clearly visible in these two projections while projection from the honeycomb plane shown the rotation of the moment about the honeycomb plane.

\begin{figure}[t]
\centering
\includegraphics[width = 8 cm]{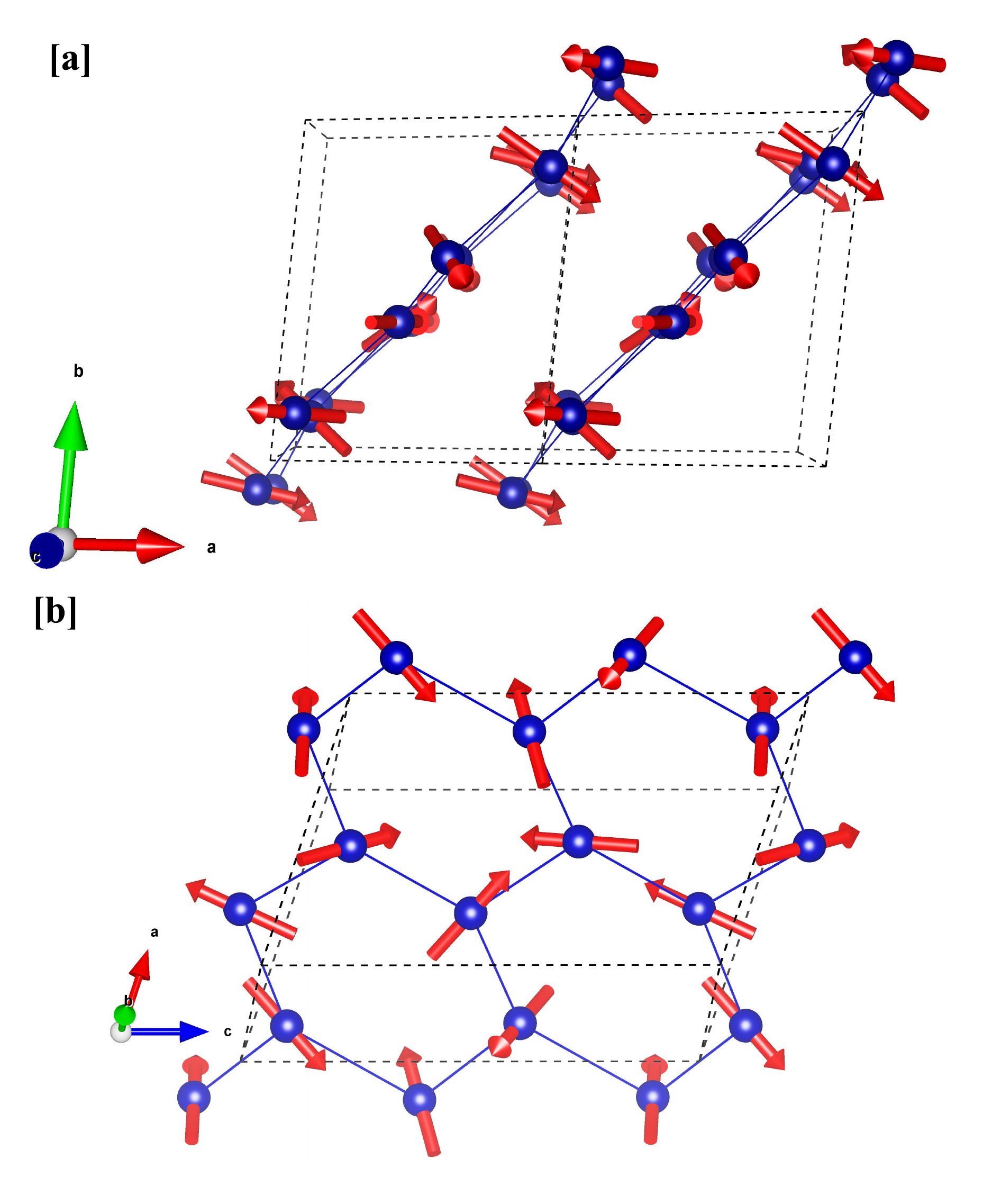}
\caption{(a) Perspective view of the spin configuration of Mn$^{2+}$ in the unit cell is shown for the incommensurate phase below $T_{N2} =$ 11.8 K. (b) depicts schematic view of spin configuration in the low-$T$ incommensurate phase obtained from the refinement of 4 K data.}
\label{COM}
\end{figure}

\section{Discussion}
$\alpha-$Mn$_2$V$_2$O$_7$ reveals two successive magnetic transitions with two distinct magnetically ordered states, $T_{N1} \approx$ 19 K and $T_{N2} \approx$ 11.8 K, from susceptibility as well as heat capacity measurements. Here, Mn$^{2+}$ ions have high-spin state and the system exhibit magnetic anisotropy due to admixture of orbital moment. Actually, $T_{N1}$ of $\approx$ 19 K is rather low compared to Curie-Weiss constant $\theta_C =-$ 34 K which indicates the presence of some frustration in the system which might be the responsible criteria to suppress the 3D magnetic ordering. The second anomaly at $T_{N2}$ was not reported previously and it is interesting to find out the nature of this transition. At first, the indication of another magnetic phase was observed from  neutron powder diffraction study (although no anomaly was found at this $T_{N2}$ in magnetization), but it was nontrivial to determine the magnetic structure solely based on powder data because of very low symmetry of $\alpha-$Mn$_2$V$_2$O$_7$ with four different Mn sites. Hence to solve these two magnetic phases single crystal has been prepared and from our careful neutron diffraction study on the single crystalline sample of Mn$_2$V$_2$O$_7$, we found commensurate magnetic structure in the intermediate state below $T_{N1}$ and incommensurate cycloidal-type structure at low temperature.

\par
Microscopically, $\alpha-$Mn$_2$V$_2$O$_7$ can be viewed as quasi-2D, with distorted honeycomb layer consisting of Mn$^{2+}$ ions parallel to $[1\bar{1}0]$ plane which is the plane along body diagonal of the unit cell. But the absence of diffuse scattering in the neutron diffraction data indicates strong interaction between the honeycomb layers. Neutron diffraction experiment have shown that this compound adopts a collinear commensurate magnetic structure below $T_{N1}$ followed by a incommensurate structure with cycloidal modulation of Mn spins at low temperature and both phases can be described by the same type of irreducible representation. The incommensurate structure has a predominant antiferromagnetic component, giving rise to satellite peaks in the vicinity of the fundamental AFM Bragg reflection. Now, the existence of this type of commensurate to incommensurate phase transition in insulators is (i) either due to competing exchange interactions, (ii) or due to relativistic effects such as spin-orbit coupling. While the first one can be found accidentally, the second mechanism depends on lattice symmetry.~\cite{roessli, vaknin} In case of $\alpha-$Mn$_2$V$_2$O$_7$, relativistic effects can be discarded as no signature was observed from our experimental data indicating the presence of spin-orbit coupling. Hence, competing exchange interactions appear to be the main driving mechanism of the incommensurate-commensurate phase transition in our case which can generate due to anisotropies in the spin Hamiltonian and/or frustration in the lattice.~\cite{bak}

\par
The reduced magnetic moment in the commensurate phase ($\approx$ 2.83 $\mu_B$) can be taken as a hallmark of quantum fluctuations triggered by the frustrated couplings of Mn spins while the moment increases rapidly below $T_{N2}$ ($\approx$ 5.3 $\mu_B$). Now, frustration in the crystal lattice is expected because of very low symmetry and 4 non-equivalent crystal sites of Mn where all positions are general. The nature of Mn - Mn interaction is superexchange via O atoms. Because of the distorted nature of Mn-honeycomb layers, there exist different types of Mn - Mn bond lengths with variable O-environment. Hence, the Mn - Mn coupling is random on a honeycomb and supposed to create competing interaction. Further we did not observe any structural incommensurability or structural transition between 280 K and 2 K. So, we can hypothesize that the incommensurate phase below $T_{N2}$ originate from sudden appearance of quantum fluctuations which is also supported from our magnetic field dependent heat capacity study. Magnetic field up to 30 kOe increases the fluctuations and supports to overcome the collinear AFM ordering and the anomaly at $T_{N2}$ shifts to the higher $T$. But the period of the spin modulation does not change with $T$ in the incommensurate regime which indicates some frozen magnetic phase below $T_{N2}$. Theoretical predictions suggest that frozen spin modulation is possible in a 2D system with random distribution of coupling.~\cite{marino}The studied compound is quasi-2D and the theoretical predictions are also in line with our hypothesis.

\section{Conclusions} 
$\alpha-$Mn$_2$V$_2$O$_7$ reveals two successive magnetic transitions at $T_{N1} =$ 19 K and $T_{N2} =$ 11.8 K indicationg two distinct magnetically ordered phases. The single crystalline sample has been grown in the floating zone furnace. No further structural chages are observed betwen 280 to  2 K in $\alpha-$Mn$_2$V$_2$O$_7$ phase. By using single crystal neutron diffraction study, we have shown that in the $T$ range 11.8 $\le T \le$ 19 K the magnetic structure is commensurate with propagation vector $k_1 = (0, 0.5, 0)$, while upon lowering temperature below $T_{N2} =$ 11.8 K a second magnetic phase towards an incommensurate ($k_2 = (0.38, 0.48, 0.5)$) 3D magnetic order emerges. The incommensurate modulation of the Mn$^{2+}$ spins of the low-$T$ phase is driven by the presence of competing exchange interactions of the next nearest neighbour Mn$^{2+}$ spins. The magnetic model at low temperature can be represented by cycloidal modulation along $ac-$plane while in the intermediate commensurate phase collinear arrangement is observed. Our study will stimulate theoretical interest and microscopic studies on the nature and origin of incommensurate to commensurate phase transition observed in frustrated insulator systems.

\par
\par
\

Neutron data were taken on the SXD diffractometer at the ISIS Neutron and Muon Source. Information on the data can be accessed through STFC ISIS Facility ~\cite{data}.

\section*{ACKNOWLEDGEMENT}
J.S. would like to thank the European Union’s Horizon 2020 research and innovation programme under the Marie Skłodowska-Curie grant agreement (GA) No 665593 awarded to the Science and Technology Facilities Council. V. Petricek likes to acknowledge the Czech Science Foundation (project no. - 18-10504S) for developing the JANA software. We thank Dr G. Stenning for help on magnetization and heat capacity measurements and ISIS Facility for providing beam time on SXD (RB1810466).

\
\
\
\


\section*{References}

\begin{thebibliography}{99}

\bibitem{meng} Z. Y. Meng, T. C Lang, S. Wessel, F. F. Assaad, and A. Muramatsu, {\it Nature} {\bf 464}, 847 (2010).
\bibitem{ladak} S. Ladak, D. E. Read, G. K. Perkins, L. F. Cohen, and W. R. Branford, {\it Nat. Phys.} {\bf 6}, 359 (2010).
\bibitem{nakatsuji} S. Nakatsuji, K. Kuga, K. Kimura, R. Satake, N. Katayama, E. Nishibori, H. Sawa, R. Ishii, M. Hagiwara, and F. Bridges, {\it Science} {\bf 336}, 559 (2012).
\bibitem{grandi} F. Grandi, F. Manghi, O. Corradini, and C. M. Bertoni, {\it Phys. Rev. B} {\bf 91}, 15112 (2015).
\bibitem{zheng} W. H. Zheng, J. Oitmaa, and C. J. Hamer, {\it Phys. Rev. B} {\bf 44}, 11869 (1991).
\bibitem{wenzel} S. Wenzel, L. Bogacz, and W. Janke, {\it Phys. Rev. Lett.} {\bf 101}, 127202 (2008).
\bibitem{tutsch} U. Tutsch, B. Wolf, S. Wessel, L. Postulka, Y. Tsui, H. O. Jeschke, I. Opahle, T. Saha-Dasgupta, R. Valenti, A. Bruhl, K. Removic-Langer, T. Kretz, H.-W. Lerner, M. Wagner, and M. Lang, {\it Nat. Commun.} {\bf 5}, 5169 (2014).
\bibitem{yan}S.M. Yan, D.A. Huse and S.R. White, {\it Science} {\bf 332}, 1173 (2011).
\bibitem{nishimoto} S. Nishimoto, V.M. Katukuri, V. Yushankhai, H. Stoll, U.K. Rößler, L. Hozoi, I. Rousochatzakis and J. van den Brink, {\it Nat. Commun.} {\bf 7}, 10273 (2015)
\bibitem{youn} Youn S J, Fischer M H, Rhim S H, Sigrist M and Agterberg D F {\it Phys. Rev. B} {\bf 85}, 220505 (2012).
\bibitem{tou} M. Touaiher, K. Rissouli, K. Benkhouja, M. Taibi, J. Aride, A. Boukhari, B. Heulin, {it Mater. Chem. Phys.} {\bf 85}, 41 (2004).
\bibitem{he1} Z. He, J.-I. Yamaura, Y. Ueda, W. Cheng, {\it Phys. Rev. B} {\bf 79}, 092404 (2009).
\bibitem{tsirlin} A. A. Tsirlin, O. Janson, H. Rosner, {\it Phys. Rev. B.} {\bf 82},  144416 (2010).
\bibitem{he2} Z. He, Y. Ueda, {\it Phys. Rev. B.} {\bf 77}, 052402 (2008) .
\bibitem{he3} Z. He, J. -I. Yamaura, Y. Ueda, W. Cheng, {\it J. Solid State Chem.} {\bf 182}, 2526 (2009).
\bibitem{calvo} C. Calvo, R. Faggiani, {\it Acta Crystallogr. B} {\bf 31},  603 (1975).
\bibitem{jhuma1} J. Sannigrahi, S. Bhowal, S. Giri, S. Majumdar, I. Dasgupta, {\it Phys. Rev. B} {\bf 91}, 220407 (R) (2015).
\bibitem{jhuma2} J. Sannigrahi, S. Giri, and S. Majumdar, {\it Solid State commun.} {\bf 228}, 10 (2016).
\bibitem{jhuma3} J. Sannigrahi, S. Giri, and S. Majumdar, {\it J. Physics and Chemistry of Solids} {\bf 101}, 1 (2017).
\bibitem{liao} J. H. Liao, F. Leroux, C. Payen, D. Guyomard and Y. Piffard, {\it J. Solid State Chem.} {\bf 121}, 214 (1996).
\bibitem{he4} Z. He, Y. Ueda and M. Itoh, {\it Solid State Commun.} {\bf 147}, 138 (2008).
\bibitem{he5} Z. He and Y. Ueda, {\it Journal of Crystal Growth} {\bf 310}, 171 (2008)
\bibitem{zhou} Z. Chuan-Cang, L. Fa-Min and D. Peng, {\it Chinese Physics B} {\bf 18}, 1674 (2009).
\bibitem{fotiev} A. A. Fotiev and L. L. Surat, {\it Russian Journal of Inorganic Chemistry} {\bf 27}, 4 (1982).
\bibitem{ueda} Z. He and Y. Ueda, {\it J. Solid State Chem.} {\bf 181}, 235 (2008).
\bibitem{keen} D. A. Keen, M. J. Gutmann, and C. C. Wilson, {\it J. Appl. Cryst.} {\bf 39}, 714 (2006).
\bibitem{SXD2001} M. Gutmann, {\it Acta Crystallogr., Sect. A} {\bf 61}, c164 (2005).
\bibitem{JANA} V. Pet\v{r}\'{i}\v{c}ek, M. Du\v{s}ek, and L. Palatinus, {\it Crystallographic computing system JANA2006: General features, Z. Kristallogr.} {\bf 229}, 345 (2014).
\bibitem{roessli} B. Roessli, J. Schefer, G. A. Petrakovskii, B. Ouladdoaf, M. Boehm, U. Staub, A. Vorotonov, and L. Bezmaternikh, {\it Phys. Rev. Lett.} {\bf 86}, 1885 (2001).
\bibitem{vaknin} D. Vaknin, J. L. Zarestky, J.-P. Rivera, and H. Schmid, {\it Phys. Rev. Lett.} {\bf 92}, 207201 (2004).
\bibitem{bak} P. Bak, {\it Rep. Prog. Phys.} {\bf 45}, 587 (1982).
\bibitem{marino} E. C. Marino, {\it Phys. Rev. B} {\bf 65}, 054418 (2002).
\bibitem{data} DOI: https://doi.org/10.5286/ISIS.E.RB1810466.




\end{thebibliography}
\end{document}